\documentclass[aps,prb,twocolumn,eqsecnum,floatfix,showpacs]{revtex4} 

\usepackage{graphicx}
\usepackage{amsmath}
\usepackage{amsfonts} 
\usepackage{bm}

\begin{document}

\title{Geodesic scattering by surface deformations of a topological insulator}
\author{J. P. Dahlhaus}
\affiliation{Instituut-Lorentz, Universiteit Leiden, P.O. Box 9506, 2300 RA Leiden, The Netherlands}
\author{C.-Y. Hou}
\affiliation{Instituut-Lorentz, Universiteit Leiden, P.O. Box 9506, 2300 RA Leiden, The Netherlands}
\author{A. R. Akhmerov}
\affiliation{Instituut-Lorentz, Universiteit Leiden, P.O. Box 9506, 2300 RA Leiden, The Netherlands}
\author{C. W. J. Beenakker}
\affiliation{Instituut-Lorentz, Universiteit Leiden, P.O. Box 9506, 2300 RA Leiden, The Netherlands}
\date{July 2010}
\begin{abstract}
We consider the classical ballistic dynamics of massless electrons on the conducting surface of a three-dimensional topological insulator, influenced by random variations of the surface height. By solving the geodesic equation and the Boltzmann equation in the limit of shallow deformations, we obtain the scattering cross section and the conductivity $\sigma$, for arbitrary anisotropic dispersion relation. At large surface electron densities $n$ this geodesic scattering mechanism (with $\sigma\propto\sqrt{n}$) is more effective at limiting the surface conductivity than electrostatic potential scattering.
\end{abstract}
\pacs{73.23.Ad, 73.25.+i, 73.50.Bk}
\maketitle

\section{Introduction}
\label{sec:intro}

Topological insulators such as ${\rm Bi}_{2}{\rm Se}_{3}$ form a new class of materials, characterized by an insulating bulk and a conducting surface.\cite{Qi10,Has10} The surface states are massless Dirac fermions with spin tied to momentum by spin-orbit coupling. Time-reversal symmetry prohibits backscattering and prevents disorder from localizing the surface states. The surface conductivity can therefore be unusually large, offering potential applications for electronics. The limitations on the conductivity of Dirac fermions imposed by random potential fluctuations are well understood (mostly from extensive studies of graphene\cite{graphene-review}). Here we study an altogether different non-electrostatic scattering mechanism, originating from random surface deformations.

The epitaxial growth of ${\rm Bi}_{2}{\rm Se}_{3}$ films is known to produce random variations in the height profile $z=\zeta(x,y)$ of the surface.\cite{Chen10} These surface deformations correspond to terraces of additional layers of the material (of typical height $H=2\,{\rm nm}$ and width $W=10\,{\rm nm}$). Since the Dirac fermions are bound to the surface, they are forced to follow its geometry. Like photons in curved space-time, the electrons follow the geodesic or shortest path between two points, although here the curvature is purely spatial.\cite{Lee09} (The metric tensor of the surface does not couple space to time.) The geodesic motion around deformations constitutes a scattering mechanism that by its very nature is energy independent, and which therefore is qualitatively different from potential scattering.

Our problem has no direct analogue in the context of graphene. Ripples of a graphene sheet do scatter the electrons, but this is not geodesic scattering: Ripples in graphene are described by gauge fields and scalar potentials in a \textit{flat} space.\cite{graphene-review} Space curvature effects may appear around conical defects (pentagon and heptagon rings), but these are rare in graphene.\cite{Vozmediano10} An early study of geodesic scattering in condensed matter that we have found in the literature is by Dugaev and Petrov,\cite{Dugaev95} with possible applications to intercalated layered crystals. The present work goes beyond their analysis by including the effects of an anisotropic dispersion relation, which is a major complication but relevant for topological insulators.

The paper is organized as follows. In Sec.~\ref{sec:scatmech} we investigate the classical motion of the surface electrons in the presence of surface deformations. The geodesic equation is solved in the regime $H/W\ll 1$ of shallow deformations, to obtain the differential scattering cross section ${\cal S}$. In Sec.~\ref{sec:conductivity} we use the linearized Boltzmann integral equation to compute the conductivity tensor $\sigma$ from ${\cal S}$. This is a notoriously difficult problem for an anisotropic dispersion relation.\cite{Zim72} In the regime $H/W \ll 1$ we are able to find a closed-form solution, by converting the integral equation into a differential equation. Results are given in Sec.~\ref{sec:results} and in Sec.~\ref{sec:expsign} we discuss the experimental signatures that distinguish geodesic scattering from potential scattering.

\section{Geodesic scattering}
\label{sec:scatmech}

\subsection{Geodesic motion}
\label{sec:geodesic}

We consider the surface of a topological insulator in the $x-y$ plane, deformed by a locally varying height $z=\zeta(x,y)$. The dispersion relation of a locally flat surface is an elliptical hyperboloid,
\begin{equation}
E=\sqrt{v_{x}^{2}p_{x}^{2}+v_{y}^{2}p_{y}^{2}+v_{z}^{2}p_{z}^{2}+\epsilon^{2}},\label{Edispersion1}
\end{equation}
where we have taken the $x,y,z$ axes as the principal axes of the elliptical cone. In general, all three velocity components $v_{x},v_{y},v_{z}$ may be different. For an isotropic dispersion relation in the $x-y$ plane we have in-plane velocities $v_{x}=v_{y}=v_{F}$, but the out-of-plane velocity $v_{z}$ may still differ. 

We have included a mass term $\epsilon$ in Eq.\ \eqref{Edispersion1} in order to have a nonzero Lagrangian,
\begin{equation}
L=\sum_{i}\dot{x}_{i}p_{i}-E
=-\epsilon\sqrt{1-\sum_{i}(\dot{x}_{i}/v_{i})^{2}},\label{Ldef1}
\end{equation}
with $\dot{x}_{i}=dx_{i}/dt=\partial E/\partial{p}_{i}$ and $i=x,y,z$. In the final equation of motion $\epsilon$ will drop out.

The constraint that the motion follows the surface implies $\dot{z}=(\partial\zeta/\partial x)\dot{x}+(\partial\zeta/\partial y)\dot{y}$, which can be used to eliminate $\dot{z}$ from the Lagrangian. The result can be written in the form
\begin{equation}
L=-\epsilon\sqrt{1-v_{x}^{-2}g_{\mu\nu}\dot{x}^{\mu}\dot{x}^{\nu}},\label{Ldef2}
\end{equation}
with $g_{\mu\nu}$ the metric tensor (made dimensionless by pulling out a factor $v_{x}^{2}$). Summation over repeated indices $\mu,\nu=1,2=x,y$ is implied and upper or lower indices distinguish contravariant or covariant vectors. 

Explicitly, we find
\begin{subequations}
\label{gzetarelation1}
\begin{align}
&g_{xx}=1+(\partial\zeta/\partial x)^{2}v_{xz}^{2},\\
&g_{yy}=v_{xy}^{2}+(\partial\zeta/\partial y)^{2}v_{xz}^{2},\\
&g_{xy}=g_{yx}=(\partial\zeta/\partial x)(\partial\zeta/\partial y)v_{xz}^{2},
\end{align}
\end{subequations}
where we have abbreviated $v_{ij}=v_{i}/v_{j}$. The inverse of the tensor $g_{\mu\nu}$, denoted by $g^{\mu\nu}$, has elements
\begin{subequations}
\label{gzetarelation2}
\begin{align}
&g^{xx}=D^{-1}[1+(\partial\zeta/\partial y)^{2}v_{yz}^{2}],\\
&g^{yy}=D^{-1}[v_{yx}^{2}+(\partial\zeta/\partial x)^{2}v_{yz}^{2}],\\
&g^{xy}=g^{yx}=-D^{-1}(\partial\zeta/\partial x)(\partial\zeta/\partial y)v_{yz}^{2},\\
&D=1+(\partial\zeta/\partial x)^{2}v_{xz}^{2}+(\partial\zeta/\partial y)^{2}v_{yz}^{2}\label{Ddef}.
\end{align}
\end{subequations}

The Euler-Lagrange equation $\partial L/\partial x^{\mu}=(d/dt)\partial L/\partial\dot{x}^{\mu}$ gives the inhomogeneous geodesic equation,\cite{Gravitation,Ber99}
\begin{equation}
\ddot{x}^{\lambda}+\Gamma^{\lambda}_{\mu\nu}\dot{x}^{\mu}\dot{x}^{\nu}=\dot{x}^{\lambda}\frac{1}{L}\frac{dL}{dt}.\label{eq:geodesics-ar}
\end{equation}
The coefficients $\Gamma^{\lambda}_{\mu \nu}$ are the Christoffel symbols,
\begin{equation}
\Gamma^{\lambda}_{\mu \nu}\equiv \frac{g^{\lambda \delta}}{2}\left(\frac{\partial }{\partial x^{\nu}}g^{\ }_{\delta \mu}+ \frac{ \partial}{\partial x^{\mu}} g^{\ }_{\delta \nu} - \frac{\partial}{\partial x^\delta} g^{\ }_{ \mu  \nu}\right). \label{Gammadef}
\end{equation}

The nonzero right-hand-side in Eq.\ \eqref{eq:geodesics-ar} may be eliminated by a reparameterization of time, from $t$ to $\tau$ such that $d\tau/dt=-L(t)/\epsilon$. We thus arrive at the homogeneous geodesic equation
\begin{equation}
\frac{d^{2}x^{\lambda}}{d\tau^{2}}+\Gamma^{\lambda}_{\mu\nu}\frac{dx^{\mu}}{d\tau}\frac{dx^{\nu}}{d\tau}=0.\label{eq:geodesics}
\end{equation}
Since $\epsilon$ does not appear in this equation of motion, it holds also in the limit of massless electrons.

\subsection{Scattering angle}
\label{sec:cross}

We consider the scattering from a surface deformation $\zeta(x,y)$ of characteristic width $W$ and height $H$ large compared to the Fermi wave length $\lambda_{F}$. The scattering may then be described by the classical equation of motion, which is the geodesic equation \eqref{eq:geodesics}.

An electron with wave vector $\bm{k}$ incident on the deformation with impact parameter $b$ at an angle $\theta_{\bm k}$ with the $x$-axis is scattered by an angle $\theta(\theta_{\bm k},b)$, resulting in a differential scattering cross section ${\cal S}(\theta_{\bm k},\theta)=|db/d\theta|$. Multiple trajectories may lead to the same scattering angle so that $\theta(\theta_{\bm k},b)$ cannot be inverted. Then the function has to be split into several invertable branches $i$ and the cross section becomes $\mathcal{S}(\theta_{\bm k},\theta)=\sum_i | d b_i(\theta_{\bm k},\theta)/d\theta|$.

These quantities may be calculated by numerically solving the geodesic equation. Analytical progress is possible in the physically relevant regime $H/W\ll 1$ of shallow deformations. As shown in App.~\ref{app:geodesic-shallow}, the scattering angle is then given by
\begin{equation}
\label{eq:scattangle}
\theta(\theta_{\bm k},b)=-\int_{-\infty}^{\infty}\tilde{\Gamma}^{y}_{xx}(\tilde{x},b)d\tilde{x}.
\end{equation}
Here $\tilde{\Gamma}^{\lambda}_{\mu\nu}(\tilde{x},\tilde{y})$ is obtained from $\Gamma^{\lambda}_{\mu\nu}(x,y)$ by a rotation of the coordinate axes over an angle $\theta_{\bm{k}}$ (so that the electron is incident parallel to the $\tilde{x}$-axis). To leading order in $H/W$ and $b/W$ the scattering angle scales as $\theta={\cal O}(H^{2}b/W^{3})$.

\begin{figure}
\includegraphics[scale=0.5,width=0.7\linewidth]{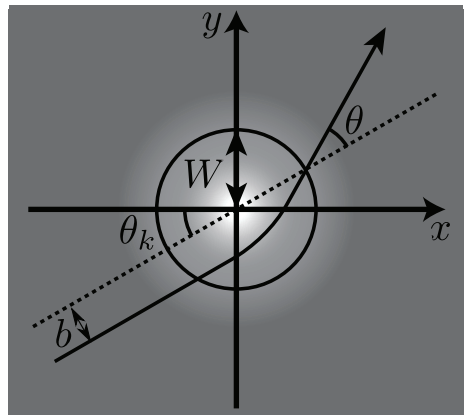}
\caption{Geodesic trajectory of an electron deflected by a circularly symmetric deformation (characteristic width $W$). The impact parameter $b$, incident angle $\theta_{\bm{k}}$, and scattering angle $\theta$ are indicated. The greyscale background shows the height profile of the Gaussian deformation \eqref{eq:Gaussian-shape}.}
\label{fig:schemeGaussian}
\end{figure}

One simple example is the case of a Gaussian deformation,
\begin{equation}
\label{eq:Gaussian-shape}
\zeta(x,y)=H \exp[-(x^2+y^2)/2 W^2],
\end{equation}
which yields (see App.~\ref{app:circularly-deformation})
\begin{equation}
\begin{split}
\theta(\theta_{\bm k},b)=& -\frac{\sqrt{\pi}}{2} \frac{H^2 v_{yz}^{\ } }{W^3} b e^{-b^2/W^2}
\\
&\times (\cos^2 \theta_{\bm{k}} +v_{yx}^2 \sin^2 \theta_{\bm{k}}) ,
\end{split}
\label{eq:scattangleGaussian}
\end{equation}
in the shallow deformation limit. The geometry is depicted in Fig.~\ref{fig:schemeGaussian}.
We will use this example throughout the paper to illustrate our general results.

\section{Calculation of the conductivity}
\label{sec:conductivity}

\subsection{Linearized Boltzmann equation}

We investigate how geodesic scattering influences the surface conductivity $\sigma$ of the topological insulator. We assume $\sigma\gg e^{2}/h$, so that we may use a semiclassical Boltzmann equation approach. In the presence of an external electric field $\bm{E}$, the occupation $f_{\bm{k}}=f_0 (E_{\bm{k}})+g_{\bm{k}}$ of the electron states deviates to first order in $\bm{E}$ according to the linearized Boltzmann equation,
\begin{equation}
\frac{\partial f_0}{\partial E_{\bm{k}}} e \bm{v}_{\bm{k}}\cdot \bm{E}= \sum_{\bm{k'}} Q(\bm{k}, \bm{k}') (g_{\bm{k}}-g_{\bm{k}'}).
\label{eq:BoltzmannEquation}
\end{equation}
Here, $\bm{v}_{\bm{k}}=\partial \mathcal{E}_{\bm{k}} / \hbar \partial \bm{k}$ is the velocity and $Q(\bm{k},\bm{k}')$ the scattering rate from $\bm{k}$ to $\bm{k}'$ (equal to $Q(\bm{k}',\bm{k})$ because of detailed balance). The sum over $\bm{k}'$ runs over all states of the ($d$-dimensional) momentum space. In the continuum limit, $\sum_{\bm{k}}\rightarrow V\int d {\bm k} / (2\pi)^{d}$, where $V$ is the $d$-dimensional volume ($d=2$ in our case). Spin degrees of freedom do not contribute to the sum since the helical surface states have definite spin direction. Particle conservation leads to the normalization condition
\begin{equation}
\label{eq:normal-condi}
\sum_{\bm{k}}g_{\bm{k}}=0.
\end{equation}

The electric field can be eliminated from Eq.~(\ref{eq:BoltzmannEquation}) by means of the vector mean free path $\Lambda_{\bm{k}}$, defined by~\cite{Zim72,Sondheimer62}
\begin{align}
&g_{\bm{k}}=\frac{\partial f_{0}}{\partial E_{\bm{k}}}e\bm{E}\cdot \bm{\Lambda}_{\bm{k}},\\
&\sum_{\bm{k}'}Q(\bm{k},\bm{k}')(\bm{\Lambda}_{\bm{k}}-\bm{\Lambda}_{\bm{k}'})=\bm{v}_{\bm{k}}.\label{eq:BoltzmannEquation2}
\end{align}
For elastic scattering, $Q(\bm{k},\bm{k'})=\delta(\mathcal{E}_{\boldsymbol{k}}-\mathcal{E}_{\boldsymbol{k}'})q(\bm{k},\bm{k'})$. Using $d\boldsymbol{k} =dk_\perp\,dS_F=d\mathcal{E}_{\bm{k}} dS_F/|\hbar \bm{v}_{\bm{k}}|$, with $dS_{F}$ a Fermi surface element, Eq.~(\ref{eq:BoltzmannEquation2}) can be rewritten in terms of the density of states $N(\cal{E}_{F}$) at the Fermi energy,
\begin{equation}
N(\mathcal{E}_{F})=(2\pi)^{-d}\oint dS_{F}\,|\hbar\bm{v}_{\bm{k}}|^{-1}.
\label{eq:DensityStates}
\end{equation}
The integral $\oint dS_{F}$ extends over the Fermi surface. The result is
\begin{equation}
V N(\mathcal{E}_{F})\langle q(\bm{k},\bm{k'}) (\bm{\Lambda}_{\bm{k}}-\bm{\Lambda}_{\bm{k}'})\rangle_{\bm{k'}}=\bm{v}_{\bm{k}},
\label{eq:BoltzmannEquation3}
\end{equation}
with $\langle \cdots \rangle_{\bm{k}}$ denoting the weighted average over the Fermi surface,
\begin{equation}
\langle f (\bm{k})\rangle_{\bm{k}}=\frac{\oint dS_{F}\,f(\bm{k})|\hbar\bm{v}_{\bm{k}}|^{-1}}{\oint dS_{F}\,|\hbar\bm{v}_{\bm{k}}|^{-1}}.
\label{eq:WeightedAverage}
\end{equation}
The normalization condition~\eqref{eq:normal-condi} becomes $\langle\Lambda_{\bm{k}}\rangle_{\bm{k}}=0$.

At zero temperature, the conductivity tensor is given by 
\begin{equation}
\label{eq:conduc-tensor-form}
\begin{split}
\bm{\sigma}&=\frac{e^{2}}{V} \sum_{\bm{k}}\delta(\mathcal{E}_{\boldsymbol{k}}-\mathcal{E}_{F})\,\bm{v}_{\bm{k}}\otimes \Lambda_{\bm{k}}
\\
&=e^{2}N(\mathcal{E}_{F})\langle\bm{v}_{\bm{k}}\otimes\Lambda_{\bm{k}}\rangle_{\bm{k}}.
\end{split}
\end{equation}
The direct product $\otimes$ indicates the dyadic tensor with elements $[\bm{v}_{\bm{k}}]_{i}[\bm{\Lambda}_{\bm{k}}]_{j}$. Substitution of Eq.~\eqref{eq:BoltzmannEquation3}  for ${\bm v}_{\bm{k}}$ and the use of $q(\bm{k},\bm{k}')=q(\bm{k}',\bm{k})$ shows that $\bm{\sigma}$ is a symmetric tensor.

For a low density $\mathcal{N}$ of scatterers, the scattering rate $q(\bm{k},\bm{k}')$ can be related to the differential cross section $\mathcal{S}$ of a single scatterer (averaged over all scatterers). In the two-dimensional case of interest here, the relation is
\begin{equation}
{\cal N}|{\bm v}_{\bm k}|{\cal S}(\theta_{\bm k},\theta_{{\bm k}'})d\theta_{{\bm k}'}= q(\bm{k},\bm{k}')\frac{V}{(2\pi)^{2}}\frac{dS_{F}'}{|\hbar\bm{v}_{\bm{k}'}|},\\
\label{eq:RelationRateToCrossSection}
\end{equation}
where $\theta_{\bm k}$ is the angle between $\bm{v}_{\bm k}$ and the $x$-axis. The Eq.~(\ref{eq:BoltzmannEquation3}) which determines the vector mean free path then takes the form
\begin{equation}
\label{eq:BoltzmannEquation4}
\mathcal{N}|\bm{v}_{\bm{k}}|\int_{0}^{2\pi} d\theta_{{\bm k}'}\,\mathcal{S}(\theta_{\bm k},\theta_{{\bm k}'})(\bm{\Lambda}_{\bm{k}}-\bm{\Lambda}_{\bm{k}'})=\bm{v}_{\bm{k}}.
\end{equation}

For the solution of this equation (and the interpretation of the results), it is convenient to follow Ziman~\cite{Zim72,Ziman61} and define an anisotropic relaxation time $\tau(\bm{k})$ by
\begin{equation}
\frac{1}{\tau(\bm{k})}=V \, N(\mathcal{E}_{F})\langle(1-\hat{v}_{\bm{k}}\cdot\hat{v}_{\bm{k}'})q(\bm{k},\bm{k}')\rangle_{\bm{k'}}.
\label{eq:tauaniso}
\end{equation}
Using Eq.\ \eqref{eq:RelationRateToCrossSection} this can be rewritten as
\begin{equation}
\frac{1}{\tau(\bm{k})}={\cal N}|\bm{v}_{\bm{k}}|\int_{0}^{2\pi} d\theta_{{\bm k}'}\,{\cal S}(\theta_{\bm k},\theta_{{\bm k}'})[1-\cos(\theta_{\bm{k}'}-\theta_{\bm{k}})].\label{eq:tauaniso-2D}
\end{equation}

\subsection{Isotropic dispersion relation}

For isotropic dispersion relations (when $E_{\bm{k}}$ depends only on $|\bm{k}|$, so that the velocity $\bm{v}=v_{F}\hat{k}$ is aligned with the wave vector), the linearized Boltzmann equation can be solved exactly.~\cite{Zim72} This applies, for example, to surfaces perpendicular to the $[111]$ direction of ${\rm Bi}_{2}{\rm Se}_{3}$. We consider this simplest case first.

Since the deformations do not have a preferred orientation and the dispersion is isotropic, the average scattering cross section ${\cal S}(\theta_{\bm{k}},\theta_{\bm{k}'})$ only depends on the scattering angle $\theta=\theta_{\bm{k}}-\theta_{\bm{k}'}$, independently of the incident direction. The solution to Eq.\ \eqref{eq:BoltzmannEquation3} is then $\bm{\Lambda}_{\bm{k}}=\tau \bm{v}_{\bm{k}}$ with a relaxation time $\tau$ given by
\begin{equation}
\frac{1}{\tau}={\cal N}v_{F}\int_{0}^{2\pi}d\theta\,{\cal S}(\theta)(1-\cos\theta).\label{tauS}
\end{equation}
Substitution into Eq.~(\ref{eq:conduc-tensor-form}) leads to a scalar conductivity $\sigma$ given by the Drude formula,
\begin{equation}
\sigma=e^{2}N(\mathcal{E}_{F})v_{F}^2 \frac{\tau}{d}=\frac{e^2 }{h}\frac{\mathcal{E}_{F}}{\hbar} \frac{\tau}{2}.
\label{eq:conductivityISO}
\end{equation}
In the second equality we inserted the density of states $N({\cal E}_{F})={\cal E}_{F}/(2\pi\hbar^{2}v_{F}^{2})$ of a Dirac cone with a circular cross section.

The regime $H/W\ll 1$ of shallow surface deformations is characterized by predominantly forward scattering ($|\theta|\ll 1$). Then the relaxation time \eqref{tauS} is given by the second moment of the scattering angle,
\begin{equation}
\frac{1}{\tau}=\tfrac{1}{2}{\cal N}v_{F}\int d\theta\,{\cal S}(\theta)\theta^{2}.
\end{equation}
We substitute the relation ${\cal S}(\theta)=\langle |d\theta(b)/db|^{-1}\rangle$, where $\langle\cdots\rangle$ indicates an average over the (randomly oriented) scatterers. The integration over scattering angles $\theta$ becomes an integration over impact parameters $b$,
\begin{equation}
\label{eq:tauFS}
\frac{1}{\tau}=\tfrac{1}{2}{\cal N}v_{F} \left\langle \int db\,\theta^{2}(b) \right\rangle.
\end{equation}

From Eq.~\eqref{eq:scattangle} we infer the scaling $1/\tau\propto W\times(H/W)^{4}$ of the relaxation rate with the characteristic height and width of the surface deformations. (The additional factor of $W$ comes from the integral over $b$.) This scaling was first obtained by Dugaev and Petrov.~\cite{Dugaev95} Eq.\ \eqref{eq:conductivityISO} then gives the scaling of the conductivity
\begin{equation}
\sigma ={\rm constant}\times\frac{e^2 }{h}\frac{ \mathcal{E}_{F} }{\hbar} \frac{1}{{\cal N}v_F }  \frac{W^3}{H^4 }.
\label{eq:WEAKconductivityISO}
\end{equation}

\subsection{Anisotropic dispersion relation}

We now turn to the case of an anisotropic dispersion relation. There is then, in general, no closed-form solution of the linearized Boltzmann equation.~\cite{Mertig99} One widely used approximation for the conductivity, due to Ziman,~\cite{Ziman61} has the form
\begin{equation}
\label{eq:Ziman-approx}
\bm{\sigma}_{\rm Ziman}=e^{2}N(\mathcal{E}_{F})\langle\bm{v}_{\bm{k}} \otimes \bm{v}_{\bm{k}}\tau(\bm{k})\rangle_{\bm{k}},
\end{equation}
with $\tau(\bm{k})$ the anisotropic relaxation time~\eqref{eq:tauaniso}. As we will show in the following, this is a poor approximation for our problem, but fortunately it is not needed: In the relevant limit $H/W\ll 1$ of scattering from shallow surface deformations an exact solution becomes possible. For shallow deformations forward scattering dominates, $|\theta|=|\theta_{\bm{k}}-\theta_{\bm{k}'}|\ll 1$. This allows for an expansion of $\bm{\Lambda}_{\bm{k}'}$ around $\theta_{\bm{k}}$, which reduces the integral equation \eqref{eq:BoltzmannEquation3} to a differential equation.

With the notation
\begin{equation}
M_{p}(\phi)=\int_{0}^{2\pi} d\theta\,{\cal S}(\phi,\phi+\theta)\theta^{p},
\end{equation}
the expansion to second order of Eq.~\eqref{eq:BoltzmannEquation4} can be written as
\begin{equation}
M_{1}(\phi)\frac{d}{d\phi}\lambda(\phi)+\tfrac{1}{2}M_{2}(\phi)\frac{d^{2}}{d\phi^{2}}\lambda(\phi)=-\frac{1}{{\cal N}}e^{i\phi}.
\label{M1M2equation}
\end{equation}
We introduced a complex variable $\lambda=\Lambda_x +i \Lambda_y$ to combine the two components of the vector mean free path. Denoting the radius of curvature of the Fermi surface by $\kappa(\phi)=dS_{F}/d\phi$, the normalization condition~\eqref{eq:normal-condi} becomes
\begin{equation}
\int_{0}^{2\pi} d\phi\,\frac{\kappa(\phi)}{v(\phi)}\lambda(\phi)=0.
\label{normalize}
\end{equation}
Once we have the solution of Eq.~(\ref{M1M2equation}), the conductivity tensor elements follow from
\begin{subequations}
\label{eq:cond-xx-yy-xy}
\begin{align}
&\sigma_{xx}\pm\sigma_{yy}=\frac{e^{2}}{h}{\rm Re}\int_{0}^{2\pi} \frac{d\phi}{2\pi}\, e^{\mp i\phi}\kappa(\phi)\lambda(\phi),\label{eq:cond1}\\
&\sigma_{xy}=\sigma_{yx}=\frac{e^{2}}{h}\tfrac{1}{2}{\rm Im}\int_{0}^{2\pi} \frac{d\phi}{2\pi}\, e^{i\phi}\kappa(\phi)\lambda(\phi)\label{eq:cond2}.
\end{align}
\end{subequations}

A further simplification is possible if the average scattering angle vanishes, $M_{1}(\phi)=0$. Then the second moment $M_{2}(\phi)$ of the scattering angle is, within the forward scattering approximation, directly related to the anisotropic relaxation time:
\begin{equation}
\label{eq:relation-aniso-relaxation-time-M2}
\frac{1}{\tau(\phi)}=\tfrac{1}{2}{\cal N}v(\phi)M_{2}(\phi).
\end{equation}
Eq.~\eqref{M1M2equation} can now be solved in terms of the Fourier transforms
\begin{subequations}
\begin{align}
&\ell_{n}=\int_{0}^{2\pi}\frac{d\phi}{2\pi}\,e^{-in\phi}v(\phi)\tau(\phi),\label{eq:ln}\\
&\kappa_{n}=\int_{0}^{2\pi}\frac{d\phi}{2\pi}\,e^{-in\phi}\kappa(\phi)\label{eq:kn},\\
&\lambda_{n}=\int_{0}^{2\pi}\frac{d\phi}{2\pi}\,e^{-in\phi}\lambda(\phi),
\end{align}
\end{subequations}
resulting in
\begin{equation}
\label{eq:lambdasol}
\lambda_{n}=\frac{\ell_{n-1}}{n^{2}}+{\rm constant}\times\delta_{n,0}.
\end{equation}
The normalization constant can be determined from Eq.\ \eqref{normalize}.

Inserting the solution into Eq.~(\ref{eq:cond-xx-yy-xy}) we obtain the conductivity
\begin{subequations}
\label{eq:cond-xx-yy-xy-1}
\begin{align}
&\sigma_{xx}\pm\sigma_{yy}=\frac{e^{2}}{h}\,{\rm Re}\sum_{n=-\infty}^{\infty} \frac{\ell_{n-1}\kappa_{-n\pm 1}}{n^{2}},\label{eq:dcond}\\
&\sigma_{xy}=\sigma_{yx}=\frac{e^{2}}{h}\tfrac{1}{2}\,{\rm Im}\sum_{n=-\infty}^{\infty} \frac{\ell_{n-1}\kappa_{-n-1}}{n^{2}}\label{eq:offdcond}.
\end{align}
\end{subequations}
For simplicity we have assumed an inversion symmetric Fermi surface, for which $\kappa_{\pm 1}=0$ so that the normalization constant in Eq.\ \eqref{eq:lambdasol} does not contribute to the conductivity.

In the case of an isotropic Fermi surface, only the Fourier components $l_{0}=v_{F}\tau$ and $\kappa_{0}=k_{F}$ are nonzero. From Eq.~\eqref{eq:cond-xx-yy-xy-1}, we then find $\sigma_{xy}=0=\sigma_{yx}$, $\sigma_{xx}=\sigma_{yy}=(e^{2}/2h)k_{F}v_{F}\tau$, in agreement with Eq.~\eqref{eq:conductivityISO}.

Comparing with the Ziman approximation \eqref{eq:Ziman-approx} for the conductivity in terms of the anisotropic relaxation time, we see that it can be written in the same form \eqref{eq:cond-xx-yy-xy-1}, but without the factor $1/n^{2}$. It therefore deviates strongly from our forward-scattering limit, except in the case of an isotropic Fermi surface (when only $n=1$ contributes).

\section{Results}
\label{sec:results}

\subsection{Isotropic dispersion relation}

In the shallow deformation limit the conductivity is given by Eq.~\eqref{eq:WEAKconductivityISO}, up to a numerical prefactor of order unity. We have calculated this prefactor for Gaussian deformations of the form \eqref{eq:Gaussian-shape}, randomly distributed over the surface. We assume that the deformations are shallow, $H/W\ll 1$. For simplicity, we also take the same parameters $H$ and $W$ for each deformation. From Eqs.~\eqref{eq:scattangleGaussian}, \eqref{eq:conductivityISO}, and \eqref{eq:tauFS} we obtain the result
\begin{equation}
\sigma^{\ }=  \frac{ 16  \sqrt{2} }{\pi \sqrt{\pi} }  \frac{\mathcal{E}_F }{\hbar v^{\ }_F \mathcal{N} } \frac{W^3}{(H v_F/v_z)^4 } \frac{e^2 }{h} .
\label{eq:GaussianISO}
\end{equation}
The factor $v_{F}/v_{z}$ is there to allow for an out-of-plane velocity $v_{z}$ that is different from the in-plane velocity $v_{x}=v_{y}=v_{F}$. The result \eqref{eq:GaussianISO} confirms the scaling behavior \eqref{eq:WEAKconductivityISO} and gives the numerical prefactor.

To relax the assumption $H/W\ll 1$ of shallow deformations, we solved the geodesic equation~(\ref{eq:geodesics}) numerically for the Gaussian case. The corresponding Christoffel symbols were taken from Eq.~\eqref{tildeGamma} with $v_{x}=v_{y}=v_{F}$. Using the scattering angle $\theta(b)$ that we obtained from the numerics, we calculated the conductivity following from Eqs.~(\ref{tauS}, \ref{eq:conductivityISO}).

\begin{figure}
\includegraphics[angle=0,scale=0.25]{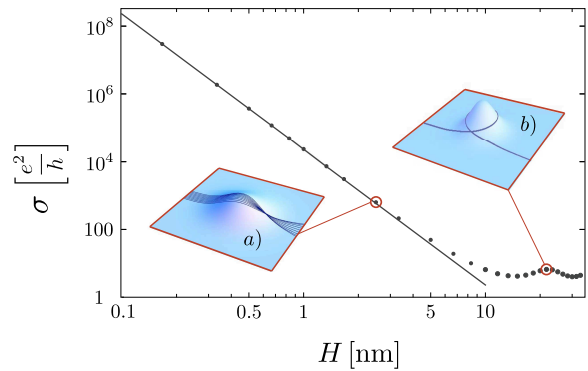}
\caption{Surface conductivity of a topological insulator as a function of the height $H$ of randomly positioned Gaussian deformations (width $W=10\,{\rm nm}$, density ${\cal N}=0.1\,W^{-2}$). We took an isotropic dispersion relation, with in-plane velocities $v_{x}=v_{y}=v_{F}=5\cdot 10^5$ m/s, and a smaller out-of-plane velocity $v_{z}=v_{F}/3$. The Fermi energy is fixed at $\mathcal{E}_{F}=150$ meV. As discussed in Sec.\ \ref{sec:expsign}, these are realistic parameter values for the $[111]$ surface of ${\rm Bi}_{2}{\rm Se}_{3}$. Dots represent numerical results whereas the line shows the shallow deformation limit~\eqref{eq:GaussianISO}.
} 
\label{fig:heightscaling}
\end{figure}

As shown in Fig.~\ref{fig:heightscaling}, the numerical results deviate from the scaling \eqref{eq:GaussianISO} only for relatively large ratios $H/W\gtrsim 0.5$. The deviations are oscillatory, due to electron trajectories that circle around the deformation  as depicted in the inset (b) of Fig.~\ref{fig:heightscaling}. Inset (a) shows generic trajectories for electrons scattering off a shallow Gaussian deformation. Notice the focussing of trajectories as an analogue of gravitational lensing. 

\subsection{Anisotropic dispersion relation}

As an example of an anisotropic dispersion relation, we consider elliptic equi-energy contours $\mathcal{E}_{\bm{k}}=\hbar(v_x^2k_x^2+v_y^2 k_y^2)^{1/2}$ with principal axes $x$ and $y$. As in the previous subsection, we investigate shallow Gaussian surface deformations. These have zero average scattering angle, $M_{1}(\phi)=0$, and second moment
\begin{equation}
\label{eq:second-moment-arbitrary}
M_2(\phi) = \frac{1}{C} (\sin^2 \phi + v^{2}_{yx} \cos^2 \phi)^2 .
\end{equation} 
The coefficient $C$ is given by
\begin{equation}
C= \frac{ 16 \sqrt{2} }{ \pi \sqrt{\pi} } \frac{W^3}{H^4 v_{y}^4/v_{z}^4}.
\end{equation}
From Eq.\ \eqref{Mpscaling} we deduce that Eq.\ \eqref{eq:second-moment-arbitrary} actually holds more generally for any circularly symmetric deformation, the only difference being in the expression for $C$.

Using Eqs.~\eqref{eq:relation-aniso-relaxation-time-M2} and \eqref{eq:ln} one obtains the Fourier coefficients
\begin{equation}
\ell_{\pm n} = \frac{C }{ \mathcal{N}}  \left(\frac{1-v^{\ }_{yx}} {1+v^{\ }_{yx}} \right)^{|n|/2} \frac{ (1+ |n| v^{\ }_{yx} +v_{yx}^2) }{v_{yx}^3}
\end{equation}
for $n$ even, and zero for $n$ odd. The elliptic dispersion relation leads to 
\begin{align}
\label{eq:kappa-phi}
&\kappa (\phi)=\frac{\mathcal{E}_F}{\hbar v_x}\, \frac{v^{\ }_{yx}}{(\sin^2 \phi + v_{yx}^2 \cos^2 \phi)^{3/2}}.
\end{align}
The Fourier coefficients $\kappa_{n}$ are also nonzero only for $n$ even. (Since their expressions are rather lengthy, we do not list them here.)

From Eq.~\eqref{eq:cond-xx-yy-xy-1} we find that the off-diagonal components of the conductivity tensor vanish, while the diagonal components are given by 
\begin{equation}
\label{eq:conductivityANISO}
\sigma_{ xx \brace yy } = \frac{e^2}{ h} \sum_{n \ge 1} \frac{1}{2n^2} (\ell_{n+1} \pm \ell_{n-1} ) (\kappa_{n+1} \pm \kappa_{n-1}).
\end{equation}
The series converges rapidly. 

\begin{figure}
\includegraphics[angle=0,scale=0.27]{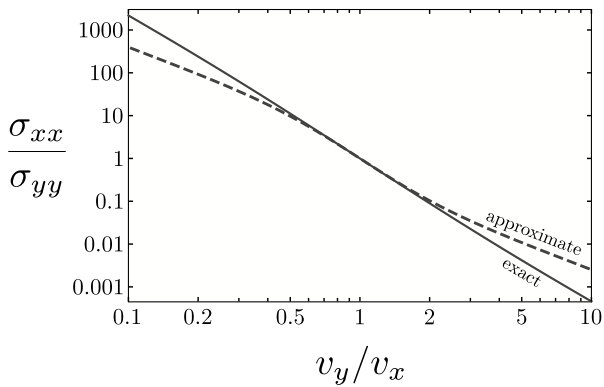}
\caption{The solid line shows the ratio of conductivities $\sigma_{xx}/\sigma_{yy}$ as a function of anisotropy $v_{y}/v_{x}$, calculated from Eq.\ \eqref{eq:conductivityANISO}. The dashed line corresponds to the Ziman approximation.} 
\label{fig:Ziman-vs-forward}
\end{figure}

The ratio $\sigma_{xx}/\sigma_{yy}$ depends only on the anisotropy $v_{yx}=v_{y}/v_{x}$. It is plotted in Fig.~\ref{fig:Ziman-vs-forward}. For comparison, we also show the Ziman approximation $\sigma^{\ }_{\rm Ziman}$ (obtained from the forward-scattering limit \eqref{eq:conductivityANISO} without the $1/n^{2}$ factor). As expected, it deviates substantially upon increasing the anisotropy (notice the logarithmic scale).

\section{Comparison with potential scattering}
\label{sec:expsign}

\subsection{Carrier density dependence}

The energy independence of the mean free path $\ell=v_{F}\tau$ is the hallmark of geodesic scattering. It implies the \textit{square root} dependence dependence $\sigma\propto\sqrt{n}$ of the conductivity on the surface electron density $n$. This follows from Eq.\ \eqref{eq:WEAKconductivityISO} with $\mathcal{E}_{F}= \hbar v_{F}\sqrt{4\pi n}$ for an isotropic Dirac cone, or more generally from the scaling $\sigma\propto S_{F}$ for a noncircular Fermi surface (of area $S_{F}\propto \sqrt{n}$). 

As discussed in the context of graphene,\cite{graphene-review,Muc10} electrostatic potential scattering typically gives a faster increase of the conductivity with increasing carrier density. Coulomb scattering from charged impurities and resonant scattering from short-range impurities both give a linear increase $\sigma\propto n$ (up to logarithmic factors). Scattering from a potential landscape with a Gaussian correlator gives an even more rapid increase $\sigma\propto n^{3/2}$. Geodesic scattering, with $\sigma\propto n^{1/2}$, would therefore form the dominant conduction-limiting scattering mechanism at high carrier densities.

\begin{figure}
\includegraphics[angle=0,scale=0.4]{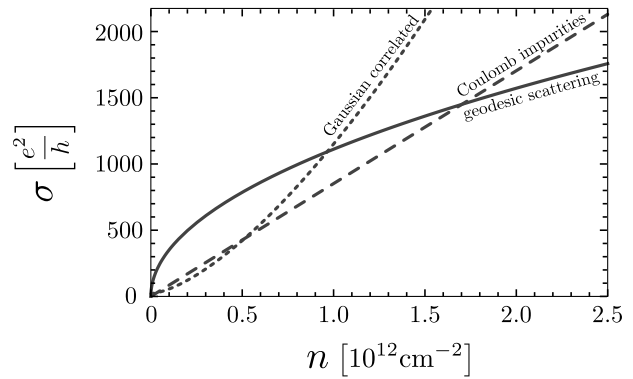}
\caption{Conductivity as a function of carrier density. The influence of three different sources of scattering is shown: surface deformations (solid line), unscreened Coulomb impurities (dashed line) and Gaussian correlated potential fluctuations (dotted line). The parameters used for the plot are given in the text.
} 
\label{fig:conductivity-n}
\end{figure}

For a quantitative comparison of geodesic and potential scattering, we consider the $[111]$ surface of ${\rm Bi}_{2}{\rm Se}_{3}$ with Gaussian deformations given by Eq.\ \eqref{eq:GaussianISO}. We take isotropic in-plane velocities $v_{x}=v_{y}=v_{F}=5\cdot 10^{5}\,{\rm m/s}$ and a smaller out-of-plane velocity $v_{z}=v_{F}/3$.~\cite{Xia09-BiSe-Exp,Zhang09-BiSe-band} We adopt the following numerical parameters for the deformations from an experimental image:\cite{Chen10} characteristic width $W= 10$ nm and height $H= 2$ nm, covering $40\% $ of the surface area so $\mathcal{N} = 10^{11}\,{\rm cm}^{-2}$. The carrier density dependence of the conductivity for geodesic scattering, following from Eq.\ \eqref{eq:GaussianISO}, is plotted in Fig.\ \ref{fig:conductivity-n} (solid curve).

To compare the geodesic scattering to typical potential scatterers, we also show the corresponding results for scattering from charged impurities (dashed) and Gaussian potential fluctuations (dotted) in Fig.\ \ref{fig:conductivity-n}.

For charged impurities (charge $Q=e$) we considered the unscreened Coulomb potential $U(\bm{r})=(Qe/4\pi\epsilon_{0}\epsilon_{r})|\bm{r}|^{-1}$, as the extreme case of a long-ranged potential. We took $\epsilon_{r}=80$ as a typical value for the dielectric constant and kept the other parameter values as before. The semiclassical conductivity is then given by~\cite{graphene-review,Nomura07}
\begin{equation}
\label{eq:CoulombISO}
\sigma=\frac{e^2 }{h}  \frac{n}{\mathcal{N}_c }\frac{2\pi\hbar^{2}v_{F}^{2}}{u_{0}^{2}},\;\;u_{0}=\frac{Qe}{4\epsilon_{0}\epsilon_{r}}.
\end{equation}
For Fig.~\ref{fig:conductivity-n} we used $\mathcal{N}_c = 2.5\times 10^{11}$ cm$^{-2}$ as the density of impurities.

For a potential landscape with Gaussian correlator (range $\xi$, dimensionless strength $U_{0}$),
\begin{equation}
\label{eq:UU-correlation}
\langle U({\bm r})  U({\bm r}') \rangle  = \frac{U_0 (\hbar v_F)^2}{ 2 \pi \xi^2} \exp \left(  -\frac{|{\bm r} - {\bm r}' |^2}{ 2\xi^2} \right),
\end{equation}
the conductivity takes the functional form~\cite{Adam09}
\begin{equation}
\label{eq:conductivity-Gaussian-potential}
\sigma =  \frac{e^2}{h} \frac{4 \pi n  \xi^2  e^{4 \pi n  \xi^2 }}{U_0 I^{\ }_1(4 \pi n  \xi^2)}.
\end{equation}
(The function $I_1$ is a Bessel function.) For Fig.\ \ref{fig:conductivity-n} we took $U_{0}=0.1$ and $\xi=W=10\,{\rm nm}$.

The parameter values used in Fig.\ \ref{fig:conductivity-n} are only for the purpose of illustration, but the point to make is that geodesic scattering dominates over potential scattering for large carrier densities.

\subsection{Anisotropy dependence of conductivity}

In the case of an anisotropic (elliptical) dispersion relation the conductivity will be direction dependent. This situation arises for example if the surface of ${\rm Bi}_{2}{\rm Se}_{3}$ is not in the $[111]$ direction. Geodesic scattering implies a certain universality for the directionality dependence of the conductivity, if we may assume that the surface deformations are shallow ($H/W\ll 1$) and without a preferential orientation (circularly symmetric on average). The ratio $\sigma_{xx}/\sigma_{yy}$ is then only a function of $v_{y}/v_{x}$, independent of other parameters (such as electron density or density and height of the deformations). This universal function is plotted in Fig.\ \ref{fig:Ziman-vs-forward} (solid curve). 

\begin{figure}
\includegraphics[angle=0,scale=0.4]{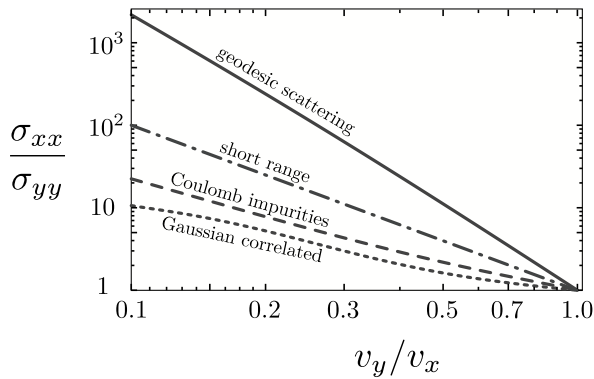}
\caption{Ratios of conductivities along the two main axes of the dispersion relation are shown as a function of anisotropy $v_{y}/v_x$. The influence of four different sources of scattering is shown: surface deformations (solid line), unscreened Coulomb impurities (dashed line), Gaussian potential fluctuations (dotted line), and short-ranged potentials (dot-dashed line). The parameters used for the plot are given in the text.} 
\label{fig:aniso}
\end{figure}

In Fig.\ \ref{fig:aniso} we compare this result for geodesic scattering with corresponding results for potential scattering. Three typical impurity potentials are considered, of different range: long-ranged unscreened Coulomb potentials, medium-ranged Gaussian potential fluctuations, and short-ranged potentials. The conductivities are obtained following the general approach of Ref.\ \onlinecite{Sorbello74}, by first computing the transition rates in Born approximation and then solving numerically the linearized Boltzmann equation. We took the same material parameter values as in the previous subsection.

The unscreened Coulomb potential gives a ratio $\sigma_{xx}/\sigma_{yy}$ which depends only on $v_{y}/v_{x}$ (dashed line). For Gaussian potential fluctuations, the ratio $\sigma_{xx}/\sigma_{yy}$ is a function of both $v_{y}/v_{x}$ and $n$. It is plotted as a dotted line in Fig.\ \ref{fig:aniso} for $n \xi^2=1$. (If $\xi=W=10$ nm this corresponds to the carrier density $n=10^{12}$ cm$^{-2}$.) In the same figure we also plot (dot-dashed line) the limit $\xi\rightarrow 0$ (at fixed $n$) of a short-ranged potential.

From the double-logarithmic plot in Fig.\ \ref{fig:aniso} one can see that there is an approximate power law dependence, $\sigma_{xx}/\sigma_{yy}\propto (v_{y}/v_{x})^{-p}$, over at least one decade. The exponent is $p\approx 3.3$ for geodesic scattering, while $p=2$ for short-range potential scattering. Scattering from long-ranged Coulomb
impurities or from medium-ranged Gaussian potential fluctuations gives $p<2$.

Anisotropic charge transport in the presence of unscreened Coulomb impurities for an elliptic dispersion relation was also discussed in the context of strained graphene.~\cite{Pereira09} There it was argued that $\sigma_{xx}/\sigma_{yy} \propto (v_y/v_x)^{-2}$ on the basis of a power-counting argument. Our numerical solution of the Boltzmann equation gives a smaller exponent $p\approx 1.3$ in that case.
 
To conclude, charge transport dominated by surface deformations has a much stronger anisotropy dependence than that governed by impurity potentials. This highly anisotropic transport behavior is a distinct characteristic of geodesic scattering.

\acknowledgments

We thank F. Hassler for valuable discussions. This research was supported by the Dutch Science Foundation NWO/FOM and by an ERC Advanced Investigator Grant.

\appendix

\section{Calculation of the scattering cross section}
\label{app:geodesic}

\subsection{Christoffel symbols in rotated basis}

In order to calculate the scattering angle in the geometry of Fig.~\ref{fig:schemeGaussian}, it is convenient to rotate the coordinate axis in the $x-y$ plane such that the electron is incident parallel to the $x$-axis. Under the linear transformation from $x,y$ to $\tilde{x}=x\cos\theta_{\bm{k}}+y\sin\theta_{\bm{k}}$, $\tilde{y}=-x\sin\theta_{\bm{k}}+y\cos\theta_{\bm{k}}$, the Christoffel symbol $\Gamma^{\lambda}_{\mu\nu}$ transforms to
\begin{equation}
\tilde\Gamma^{\lambda}_{\mu\nu}(\tilde{x},\tilde{y})=\frac{\partial \tilde{x}^{\lambda}}{\partial x^{\lambda'}}\Gamma^{\lambda'}_{\mu'\nu'}(x,y)\frac{\partial x^{\mu'}}{\partial \tilde{x}^{\mu}}\frac{\partial x^{\nu'}}{\partial \tilde{x}^{\nu}}.\label{Gammatildedef}
\end{equation}

Using the expressions \eqref{gzetarelation1}, \eqref{gzetarelation2}, \eqref{Gammadef} for metric tensor and Christoffel symbols, we arrive at
\begin{widetext}
\begin{subequations}
\label{tildeGamma}
\begin{align}
&\tilde{\Gamma}^{x}_{\mu\nu}=D^{-1}\frac{\partial^{2}\zeta}{\partial\tilde{x}^{\mu}\partial\tilde{x}^{\nu}}\left[v_{xz}^{2}\frac{\partial\zeta}{\partial\tilde{x}}-(v_{xz}^{2}-v_{yz}^{2})\sin\theta_{\bm{k}}\left(\frac{\partial\zeta}{\partial\tilde{x}}\sin\theta_{\bm{k}}+\frac{\partial\zeta}{\partial\tilde{y}}\cos\theta_{\bm{k}}\right)\right],\label{tildeGamma1}\\
&\tilde{\Gamma}^{y}_{\mu\nu}=D^{-1}\frac{\partial^{2}\zeta}{\partial\tilde{x}^{\mu}\partial\tilde{x}^{\nu}}\left[v_{yz}^{2}\frac{\partial\zeta}{\partial\tilde{y}}-(v_{xz}^{2}-v_{yz}^{2})\sin\theta_{\bm{k}}\left(\frac{\partial\zeta}{\partial\tilde{x}}\cos\theta_{\bm{k}}-\frac{\partial\zeta}{\partial\tilde{y}}\sin\theta_{\bm{k}}\right)\right].\label{tildeGamma2}
\end{align}
\end{subequations}
The factor $D$ from Eq.\ \eqref{Ddef}, written in terms of the rotated coordinates, reads
\begin{equation}
D=1+v_{xz}^{2}\left(\frac{\partial\zeta}{\partial\tilde{x}}\cos \theta_{\bm{k}}-\frac{\partial\zeta}{\partial\tilde{y}}\sin \theta_{\bm{k}}\right)^{2} +v_{yz}^{2}\left(\frac{\partial\zeta}{\partial\tilde{x}}\sin \theta_{\bm{k}}+\frac{\partial\zeta}{\partial\tilde{y}}\cos \theta_{\bm{k}}\right)^{2}.\label{Ddef2}
\end{equation}
\end{widetext}

The Christoffel symbols \eqref{tildeGamma} appear in the geodesic equation for the rotated coordinates,
\begin{equation}
\frac{d^{2}\tilde{x}^{\lambda}}{d\tau^{2}}+\tilde{\Gamma}^{\lambda}_{\mu\nu}\frac{d\tilde{x}^{\mu}}{d\tau}\frac{d\tilde{x}^{\nu}}{d\tau}=0.\label{eq:geodesics3}
\end{equation}

\subsection{Geodesic equation for shallow deformation}
\label{app:geodesic-shallow}

The geodesic equation \eqref{eq:geodesics3} can be considerably simplified in the shallow deformation limit $H/W\ll 1$. Let us consider a particle incident on a deformation along the $\tilde{x}$-direction from $-\infty$ with impact parameter $b$ and velocity
\begin{equation}
v=v_{x}v_{y}(v_y^2 \cos^2 \theta_{\bm{k}} + v_x^2 \sin^2 \theta_{\bm{k}})^{-1/2}.
\end{equation}
Since the derivative $d\tilde{y}/d\tau$ is smaller than $d\tilde{x}/d\tau$ by a factor $(H/W)^{2}$, we can drop this derivative from the geodesic equation. The result is
\begin{subequations}
\begin{align}
\frac{ d^2 \tilde{x} }{d \tau^2}+ \tilde{\Gamma}^x_{xx} \left(\frac{d \tilde{x}}{d\tau}\right)^2&=0,\\
\frac{ d^2 \tilde{y} }{d \tau^2}+\tilde{\Gamma}^y_{xx} \left(\frac{d \tilde{x}}{d\tau}\right)^2&=0.
\end{align}
\end{subequations}
Furthermore, since $d\tilde{x}/d\tau=v[1+{\cal O}(H/W)^{2}]$, we can write $d/d\tau=vd/d\tilde{x}$. This leads to
\begin{equation}
\frac{d^2 \tilde{y}}{d\tilde{x}^2}=-\tilde{\Gamma}^y_{xx}.
\end{equation}

The scattering angle $\theta\ll 1$ is obtained from $\theta=\lim_{\tilde{x}\rightarrow\infty}d\tilde{y}/d\tilde{x}$, hence
\begin{equation}
\label{eq:theta-b-approxi-app}
\theta(\theta_{\bm{k}},b) = - \int_{-\infty}^{\infty}\left. \tilde{\Gamma}^y_{xx} \, d\tilde{x}\,\right|_{\tilde{y}\rightarrow b}.
\end{equation}
Inserting Eq.~\eqref{tildeGamma2} into Eq.~\eqref{eq:theta-b-approxi-app} and noting that $D=1+{\cal O}(H/W)^{2}$, we obtain the scattering angle to leading order in $H/W$, 
\begin{align}
\label{eq:theta-b-approxi-app-1}
\theta(\theta_{\bm{k}},b) &= - \int_{-\infty}^{\infty}  d\tilde{x}\, \left[\left( \alpha \frac{\partial \zeta}{\partial \tilde{y}}-\gamma \frac{\partial \zeta}{\partial \tilde{x}} \right)  \frac{\partial^2 \zeta}{\partial \tilde{x}^2}\right]_{\tilde{y}\rightarrow b}.
\end{align}
We abbreviated
\begin{subequations}
\begin{align}
\alpha&=v_{yz}^2 \cos^2 \theta_{\bm k} +v_{xz}^2 \sin^2 \theta_{\bm k}, \\
\gamma&=(v_{xz}^2-v_{yz}^2) \sin \theta_{\bm k} \cos \theta_{\bm k}.
\end{align}
\label{eq:parameters}
\end{subequations}

\subsection{Circularly symmetric deformation}
\label{app:circularly-deformation}

For a circularly symmetric height profile $\zeta(x,y)$, dependent only on $r=\sqrt{x^{2}+y^{2}}=\sqrt{\tilde{x}^{2}+\tilde{y}^{2}}$, the term proportional to $\gamma$ in Eq.~\eqref{eq:theta-b-approxi-app-1} vanishes (because it is an integral over an odd function of $\tilde{x}$). The expression for the scattering angle thus simplifies further to
\begin{align}
\label{eq:theta-b-approxi-circularly}
\theta(\theta_{\bm{k}},b) = -  \alpha  \int_{-\infty}^{\infty}  dx\,\left[ \frac{\partial \zeta}{\partial y} \,\frac{\partial^2 \zeta}{\partial x^2}\right]_{ y\rightarrow b}.
\end{align}
For the Gaussian deformation \eqref{eq:Gaussian-shape} we obtain the scattering angle \eqref{eq:scattangleGaussian} given in the main text.

The entire dependence of the scattering angle $\theta$ on the angle of incidence $\theta_{\bm{k}}$ is contained in the prefactor $\alpha$. This implies that the moments $M_{p}=\int db\,\theta^{p}$ of the scattering angle depend on the angle of incidence as
\begin{equation}
M_{p}(\theta_{\bm{k}})=c_{p}\alpha^{p}=c_{p} v_{xz}^p (\sin^{2}\theta_{\bm{k}}+ v_{yx}^{2}\cos^{2}\theta_{\bm{k}} )^{p},\label{Mpscaling}
\end{equation}
with $c_{p}$ a coefficient independent of $\theta_{\bm{k}}$.

\end{document}